\documentclass[journal]{IEEEtran}
\usepackage{epstopdf}
\usepackage{cite}
\usepackage{amsmath,amssymb,amsfonts}
\usepackage{algorithmic}
\usepackage{graphicx}
\usepackage{textcomp}
\usepackage{amsfonts}
\usepackage{mathrsfs} 
\usepackage{epsfig}
\usepackage{subfigure}
\usepackage{algorithm}
\usepackage{caption2}
\usepackage{enumerate}
\usepackage{multirow}
\usepackage{array}
\usepackage{bm}
\usepackage{amsmath}
\usepackage{slashbox}
\usepackage{bbm}
\usepackage{color}
\usepackage{stfloats}
\usepackage[table]{xcolor}
\usepackage{colortbl}
\usepackage{makecell}

\newcommand{\PreserveBackslash}[1]{\let\temp=\\#1\let\\=\temp}
\newcolumntype{I}{!{\vrule width 1pt}}
\newcolumntype{C}[1]{>{\PreserveBackslash\centering}p{#1}}
\newcolumntype{R}[1]{>{\PreserveBackslash\raggedleft}p{#1}}
\newcolumntype{L}[1]{>{\PreserveBackslash\raggedright}p{#1}}

\newcounter{algleo}
\newlength{\lefttab}
\newlength{\numberoffset}
 {\trivlist
  \topsep=0pt\parsep=0pt\itemsep=0pt
  \addtolength{\lefttab}{1.25em}
  \addtolength{\numberoffset}{1.25em}
  \leftskip=\lefttab}%
 {\endtrivlist}

\makeatletter
\def\hlinew#1{%
  \noalign{\ifnum0=`}\fi\hrule \@height #1 \futurelet
   \reserved@a\@xhline}
\makeatother

\newtheorem{remark}{Remark}
\renewcommand{\baselinestretch}{1}
\def\BibTeX{{\rm B\kern-.05em{\sc i\kern-.025em b}\kern-.08em
    T\kern-.1667em\lower.7ex\hbox{E}\kern-.125emX}}

\allowdisplaybreaks[4]

\begin{document}
\title{Composite Disturbance Filtering: A Novel State Estimation Scheme for Systems With Multi-Source, Heterogeneous, and Isomeric Disturbances
%
\thanks{This work was supported in part by the National Natural Science Foundation of China under Grants 62003016, 61751302, 62073018, 62122007, 62103013, 61627810, 61603021; and in part by the National Key Research and Development Program of China under Grant 2020YFA0711200. {\it (corresponding author: Wenshuo Li and Zidong Wang.)}}}

\author{Lei Guo, Wenshuo Li, Yukai~Zhu,~Xiang Yu,~Zidong~Wang
\thanks{L.~Guo is with the School of Automation Science and Electrical Engineering, Beihang University, Beijing, 100191 China. Email: {\tt lguo@buaa.edu.cn}}
\thanks{W.~Li is with Hangzhou Innovation Institute, Beihang University, Hangzhou, 310051 China. Email: {\tt wslibuaa@126.com}}
\thanks{Y.~Zhu is with Department of Automation Science and Electrical Engineering, Beihang University, Beijing 100191, China}
\thanks{X.~Yu is with Department of Automation Science and Electrical Engineering, Beihang University, Beijing 100191, China}
\thanks{Zidong~Wang is with Department of Computer Science, Brunel University London, Uxbridge UB8 3PH, U.K. Email: {\tt Zidong.Wang@brunel.ac.uk}}}


\markboth{{\it submitted}} {Guo: Using the style file IEEEtran.sty} \maketitle

\begin{abstract}
State estimation has long been a fundamental problem in signal processing and control areas. The main challenge is to design filters with ability to reject or attenuate various disturbances. With the arrival of big data era, the disturbances of complicated systems are physically multi-source, mathematically heterogenous, affecting the system dynamics via isomeric (additive, multiplicative and recessive) channels, and deeply coupled with each other. In traditional filtering schemes, the multi-source heterogenous disturbances are usually simplified as a lumped one so that the ``single'' disturbance can be either rejected or attenuated. Since the pioneering work in 2012, a novel state estimation methodology called {\it composite disturbance filtering} (CDF) has been proposed, which deals with the multi-source, heterogenous, and isomeric disturbances based on their specific characteristics. With the CDF, enhanced anti-disturbance capability can be achieved via refined quantification, effective separation, and simultaneous rejection and attenuation of the disturbances. In this paper, an overview of the CDF scheme is provided, which includes the basic principle, general design procedure, application scenarios (e.g. alignment, localization and navigation), and future research directions. In summary, it is expected that the CDF offers an effective tool for state estimation, especially in the presence of multi-source heterogeneous disturbances.
\end{abstract}


\section{Motivation and Background}\label{sec:1}
State estimation is a fundamental issue in a wide range of applications including integrated navigation, remote sensing, target tracking, and cyber-physical systems \cite{Jazwinski1970,Anderson1979,ChenGuo1991BOOK,Shalom2001,MurrayAstrom2003,TII2016NCS,TII2021CPS}. It has been recognized that state estimation also plays a key role in control and even AI fields, as the next generation of AI largely relies on estimation and perception of the motion information \cite{NeuroAIWhitepaper}. With the arrival of big-data and big-model era, practical systems and information flows have become ever more complicated. In the general case, the complicated systems inevitably suffer from disturbances that are physically multi-source, mathematically heterogenous, affecting the system via isomeric (additive, multiplicative and recessive) channels, and deeply coupled with each other. Therefore, state estimation in the presence of multi-source, heterogeneous, and isomeric disturbances is an important research topic.

The celebrated Kalman filtering (KF), established in the 1960's, is arguably the most widely used approach due to its optimality in the linear minimum mean-square error sense, simplicity in structure, and ease of implementation \cite{Kalman1960, Kalman1961}. Following the Bayesian principle, primitive results on the KF have mainly concentrated on the systems with Gaussian noises. However, even in view of stochastic processes, the optimality of the classical KF can no longer be guaranteed for systems with nonlinear dynamics or non-Gaussian noises \cite{Sorenson1970,Guo2010SDCBOOK,Ding2018}. Several variants of the KF have been developed to address the nonlinear state estimation problem. Among them, the extended KF (EKF) utilizes Taylor series expansion to transform the nonlinear system model into an approximate linear one, and the common KF procedure is applied \cite{Jazwinski1970, Schmidt1981}. The unscented KF (UKF) exploits the unscented transform to calculate the mean and covariance of the state variable \cite{Julier2000}. For the quadrature KF and cubature KF (CKF), the Gauss-Hermite quadrature rule and the spherical-radial cubature rule are, respectively, employed to calculate the Gaussian weighted integrals \cite{Ito2000, Arasaratnam2007, Arasaratnam2009, Qiu2020}. Moreover, several robust KF \cite{Xie2007TCAS,Zhang2012, XiaYQ2020BOOK} and adaptive KF \cite{Chaer1997, Sarkka2009TAC, YHuang2018} algorithms have been proposed to effectively deal with the inaccurate noise statistics.

When it comes to the systems with strong non-Gaussianity, the entire probability density function (PDF) is required to be tracked with the filter. To this end, two representative schemes have been developed respectively: the Monte-Carlo-sampling-based filter and the stochastic distribution filter (SDF). The ensemble KF (EnKF) and particle filter (PF) both utilize random samples to globally approximate the posterior PDF, thereby belonging to the former \cite{Gordon1993,Arulampalam2002,Cappe2007,Ozkan2014}. The difference between them is that the EnKF employs a linear shift to propagate the random samples \cite{EnKF01,EnKF02,Evensen2009} while the PF uses reweighting \cite{LiTiancheng2015,WLi2019Auto,WLi2020Auto}. On the other hand, the SDF scheme, first established in the 2000's, is capable of achieving non-Gaussian state estimation via PDF shape approximation and error entropy minimization \cite{WangH2000Book,Guo2005TRAS,Guo2005Auto,Guo2010SDCBOOK,Tian2017SCIS}. Recently, the B-spline expansion has been used in \cite{Guo2006TSP, TLi2009} to transform the non-Gaussian stochastic system into a deterministic one, based on which the non-Gaussian disturbances are attenuated. In \cite{Guo2009TAC}, an entropy optimization filtering scheme has been proposed, where the target signal and the non-Gaussian noise are separated according to the different ways they affect the output signal. Meanwhile, the SDF has been successfully applied in many industrial fields such as the paper-making and chemical processes \cite{APPSDF01,APPSDF02}.

The above-mentioned schemes are mainly applicable to the stochastic disturbances and customarily referred to as the {\it stochastic filters}. Nevertheless, sometimes it is difficult to obtain the statistical information of disturbances, whereas the upper bound on the norm of the disturbance can be quantified. Targeting at the norm-bounded disturbances, the robust design tool (e.g. $H_{\infty}$ technique) has been employed for filter design without noise statistics \cite{Yaesh1992,Lewis2007BOOK,Ding2018}. For example, the mixed $H_{2}/H_{\infty}$ filtering problem has been studied in \cite{Guo2006CSSP,Gao2005}, where both the norm-bounded and stochastic disturbances are taken into account. In \cite{Wang2006TSP}, the norm-bounded disturbance has been considered jointly with time-delays and packet dropouts, and a robust $H_{\infty}$ filter has been designed to ensure the prescribed disturbance attenuation performance. Subsequently, this elegant result has been extended in \cite{Dong2010TSP1} to deal with both stochastic and norm-bounded disturbances and in \cite{Dong2010TSP2} to handle variance constraints, uncertain coefficient matrices, and multiple missing measurements. The $H_{\infty}$ technique has been incorporated into the sequential particle generation procedure in \cite{Miguez2004,Yu2013} to design a robust filter in the absence of noise statistics. The statistical reference in the conventional PF is replaced by a user-defined cost function that measures the quality of state estimate.

In addition to the statistical and norm-bound information utilized in the aforementioned stochastic and robust filtering methodologies, the mass data that come with the big-data era has made it possible to extract dynamic features of the disturbances. With various time series analysis tools, the disturbances can usually be described as time-varying signals with partially known dynamical information, which is especially true for the motion systems. Therefore, a natural idea is to exploit such dynamic characteristics to construct a disturbance observer (DO) for real-time disturbance rejection. The primitive results on DO have mainly focused on the frequency-domain design \cite{Ohishi1987_DO}, while the corresponding time-domain design has been restated in \cite{Chen2004_NDO} for a class of nonlinear systems. Up to now, the DO has become a popular method to enhance anti-disturbance capability in the fields of control theory and engineering \cite{Chen2016_DOBSurvey}. In \cite{Guo2005_NDO}, the DO-based composite controller has been firstly designed to achieve simultaneous disturbance rejection and attenuation, and a linear matrix inequality (LMI) condition has been rigorously established for the closed-loop stability. In order to overcome the dependence on the accurate dynamic model, a robust DO has been developed in \cite{Wei2010_RDO}, where the dynamic uncertainties are treated as a norm-bounded variable. Notice that most reported results relevant to DO have been concerned with the control issues, and the corresponding state estimation problems have not been thoroughly investigated.

In most existing literature on estimation problems, the disturbances have been merged as a single-type one. Such simplification may lead to unsatisfactory estimation performance as the specific characteristics of disturbances have not been sufficiently exploited. In fact, the complicated systems in practical engineering are always subject to multi-source heterogeneous disturbances, which highlights the need for refined anti-disturbance estimation methods. In this paper, we will give three examples in which the multi-source heterogeneous disturbances are deeply coupled: initial alignment, indoor localization, and integrated navigation systems. A representative example is the rotor unmanned aerial vehicle (UAV) indoor localization, where the information of ultra-wide-band (UWB) and UAV dynamics needs to be fused. As shown in Section~\ref{sec:2-1}, the disturbances in the localization system are physically multi-source, mathematically heterogenous, and entering the localization model in different ways. If not properly tackled, such disturbances would cause severe degradation to the localization performance. In this sense, there is an urgent need to develop refined filtering scheme by fully exploiting the specific disturbance characteristics.

Fortunately, the refined control problems under multi-source heterogeneous disturbances have been tackled via a novel methodology called composite hierarchical anti-disturbance control (CHADC). Motivated by the pioneering work in \cite{Guo2005_NDO}, the theoretical framework of CHADC has been established and developed since the early 2000's \cite{Guo2014BOOK, Guo_Cao2014survey}. Generally speaking, the CHADC is based on explicit characterization and treatment of the heterogeneous disturbances, and can be realized with a composite ``X-DO plus Y-controller'' structure \cite{Guo2005_NDO,Wei2010_RDO}. According to specific properties of the disturbances, the ``X-DO'' can be designed as standard linear DO \cite{Guo2005_NDO} or other types of DOs including robust DO \cite{Wei2010_RDO}, adaptive DO \cite{Guo_Wen2011_ADO, LiGuo2022_APDO}, adaptive sliding-mode DO \cite{Zhu2019_ASMDO}, adaptive switching DO \cite{Zhu2019_SwitchingDO}, iterative learning DO \cite{Maeda2015_ILDO}, fuzzy DO \cite{Wu2014_FuzzyDO}, neural network (NN) based DO \cite{Sun2017_NNDO}, fixed-time DO \cite{Zhou2021_FTDO}, and predictive DO \cite{Wen2020_DP} according to different characteristics of the disturbances. Meanwhile, the ``Y-controller'' can be PID, robust, adaptive, sliding-mode, ADRC, intelligent, or other advanced controllers based on various performance objectives, see \cite{Guo_Bluebook,Guo2022CJA} for examples. The CHADC can enhance the anti-disturbance capability via three key tools including refined quantification, separability analysis, and simultaneous compensation and attenuation of the multi-source heterogenous disturbances. So far, CHADC has been successfully applied to a broad class of practical systems (see Table.~\ref{Tab1} for typical application cases).

The CHADC is a refined anti-disturbance control method, where the separation of multiple disturbances (rather than the estimation of single disturbance using the extended state observer (ESO) \cite{LiSH2012TIE,Xue2020} or unknown input observer (UIO) \cite{Darouach1997,Darouach2003}), disturbance controllability (rather than state controllability), and adaptive variance principle (rather than invariance principle) have been addressed \cite{Guo_Cao2014survey,Guo_Cao2012,Guo2020SSI,Guo2020CJA,Guo_Bluebook,Guo2022CJA}.
Following the developments of the CHADC theory, the safety, green and immune control problems have been addressed for unmanned systems under disturbance and confrontation environments, where the three anti-disturbance control layers of methodology, system and behaviour are proposed \cite{Guo_Bluebook, Guo2020CJA, Guo2022CJA}. More recently, an optimal reconstruction scheme has been proposed in \cite{Reconstruction01} based on performance degradation analysis in the presence of disturbances. A disturbance utilization scheme has been presented in \cite{Jia2022RAL} where the estimated aerodynamic drag is used as a damping term in the flight control law of rotor UAVs. In \cite{Cui2023IET}, a ``green'' anti-disturbance control scheme with maximum velocity constraints is developed for gimbal servo systems.
\begin{table}[ht]
\caption{Typical applications of the CHADC}
  \label{Tab1}
  \centering
  \fontsize{10}{15}\selectfont
\begin{tabular}{C{4.4cm}IC{2.8cm}}
\hline
Plant & Reference   \\
\rowcolor{gray!10}
\hline
Flexible satellites   & \cite{LiuH2012,Zhu2019CEP}    \\
\hline
Hypersonic vehicles   & \cite{Yang2013,WangN2014ND}   \\
\rowcolor{gray!10}
\hline
Rotor UAVs            & \cite{GuoKX2020CEP,Jia2022TAES}  \\
\hline
Robot manipulators    & \cite{Zhu2019_ASMDO,Zhu2019_SwitchingDO} \\
\rowcolor{gray!10}
\hline
Servo systems         &  \cite{Yan2018,Cui2022TIE,Cui2023TAES}   \\
\hline
Power electronic devices         &  \cite{Yang2015,Durra2019TIE}   \\
\rowcolor{gray!10}
\hline
\end{tabular}
\end{table}

\begin{figure*}[ht]
\centering
\includegraphics[width=0.72\textwidth]{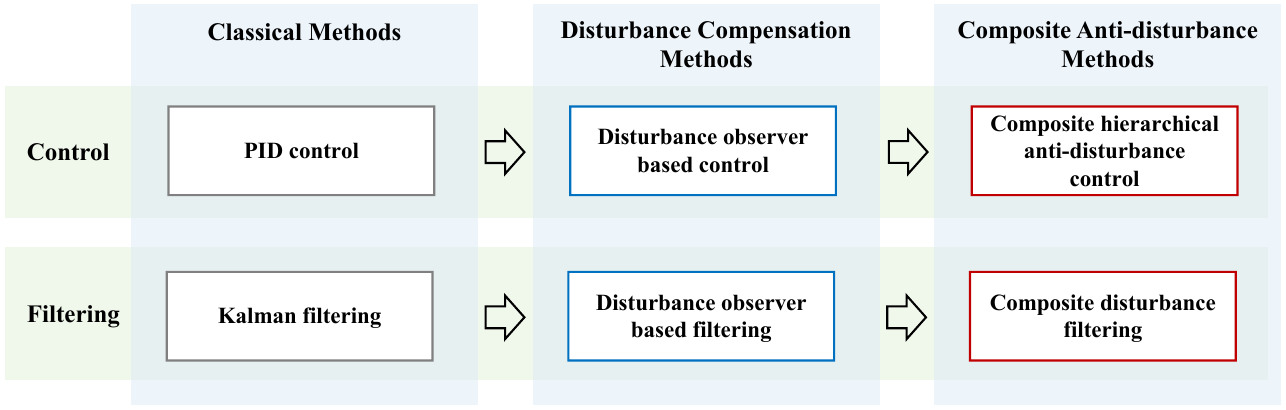}
\caption{Analogy between control and filter design.}
\label{Fig00}
\end{figure*}

The composite anti-disturbance strategy in CHADC can also be used to deal with estimation problems. Following this idea, the composite disturbance filtering (CDF) scheme has been put forward to address the state estimation problem for systems with multi-source heterogeneous disturbances \cite{Guo2014BOOK}. The analogy between the control and filter design has been illustrated in Fig.~\ref{Fig00}. Specifically, the CDF is a refined filtering methodology which aims at simultaneous disturbance rejection, attenuation and absorption via a DO-based composite hierarchical anti-disturbance estimation (CHADE) architecture. Up to now, a series of composite filter structures have been proposed within the CDF framework, which include composite DO+robust filter, composite DO+Kalman filter, composite DO+SDF, and composite DO+PF. It should also be pointed out that the hierarchical disturbance compensation and state stabilization design in the CHADC has been replaced in the CDF by refined error quantification and optimization for joint state and disturbance estimation \cite{Cao2009CDC,Guo_Cao2012,Cao_Guo2012,WLi2021TSMCA,WLi2021JFI,WLi2022TIM,LiGuo2022_APDO}.

In comparison with the existing methods, the distinctive features of the CDF scheme can be summarized as follows:
\begin{itemize}
  \item[i)] Different from the existing filtering schemes which can either reject (e.g. the disturbance observer based filtering \cite{Darouach1997,Darouach2003}) or attenuate (e.g. the KF \cite{Kalman1960} and $H_{\infty}$ filtering \cite{Yaesh1992}) a single-type disturbance, the CDF is capable of simultaneous rejection and attenuation of heterogeneous disturbances via a well-designed composite filter structure;
  \item[ii)] Based on the refined quantification of multi-source heterogeneous disturbances, the CDF can realize effective separation of the deeply coupled disturbance signals, which is especially desirable in signal identification and abnormity diagnosis;
  \item[iii)] By resorting to the generalized observability analysis and explicit error quantification rather than the existing methods for brute-force disturbance rejection, the CDF enables disturbance utilization, optimal reconstruction and performance redesign.
\end{itemize}

In this paper, advances and prospects the CDF will be discussed. The rest of this paper is organized as follows. Refined quantification and separability analysis of the multi-source heterogeneous disturbances are discussed in Section~\ref{sec:2}. The general design framework and typical realizations (composite DO+robust filter and composite DO+stochastic filter) of the CDF are elaborated in Section~\ref{sec:3}. In Section~\ref{sec:4}, the effectiveness of the CDF is illustrated via application results in initial alignment of inertial navigation systems (INSs), indoor localization of rotor UAVs, and skylight polarization aided integrated attitude determination. Conclusions and future research directions are provided in Section~\ref{sec:5}.

\section{Composite disturbance filtering: Modeling and analysis}\label{sec:2}
The major advantage of CDF is that the characteristics of disturbances can be sufficiently utilized. Hence, a prerequisite of CDF design is refined quantification and separability analysis of the multi-source, heterogeneous, and isomeric disturbances which are deeply coupled. In this section, the modeling and analysis of systems with multi-source, heterogeneous, and isomeric disturbances will be introduced.

\subsection{Deeply coupled modeling and refined quantification}\label{sec:2-1}
In the age of big data and big models, practical systems and information flows become extremely complicated. It is quite common that the disturbances affecting a practical system are physically multi-source, mathematically heterogenous, affecting the system dynamics via isomeric (additive, multiplicative and recessive) channels. Furthermore, there may exist complex interactions between different disturbance signals and between the system state and the disturbances. Hence, novel modeling and analysis methods is required to deal with the resultant deeply coupled disturbances.

For ease of illustration, let us take the UWB-based localization problem for rotor UAV working in narrow spaces as an example. It will be shown that many complex systems can be handled in the similar way. Typically, the data from UWB sensors needs to be fused with the UAV dynamics so as to generate more accurate localization results \cite{Barbieri2021,Jia2022TIM}. Hence, the fusion algorithm should be able to handle both UAV dynamic uncertainties and UWB measurement errors.

As described in \cite{Jia2022TIM}, the localization system of rotor UAVs is subject to multi-source disturbances. Specifically, the aerodynamic drag force induced by external wind, which can be a prominent issue in narrow spaces where the proximity effect is significant. Besides, the loss of efficiency (LoE) of the motor, arising from damage or aging of the motor components, constitutes another factor that leads to model uncertainty. In the measurement equation, the UWB noises are skew-$t$ instead of Gaussian due to non-line-of-sight propagation and multi-path effects. Furthermore, the interactions among different sources of disturbances cannot be ignored. For example, the aerodynamic drag force depends on the attitude and position of the UAV, which is accordingly related to LoE of the motor. Conversely, when the external wind becomes larger, greater motor torque is required to maintain stable flight, accelerating the aging process of the device and inducing a more severe LoE. In view of the aforementioned factors, there is no doubt that the localization problem is essentially a state estimation problem with multi-source, deeply coupled disturbances.

The localization model is a special case of the following generic one where multiple disturbances coexist:
\begin{equation}
\label{General01}
\left\{\begin{aligned}
E\dot{x}=&f_{\Delta}(x,\Xi u)+B\omega_{0}+B_{1}\omega_{1}+B_{2}\omega_{2}\\
y=&h_{\Delta}(x)+D\omega_{0}+D_{1}\omega_{1}+D_{2}\omega_{2}\\
\dot{\omega}_{0}=&g_{\Delta}(x,\Xi u)\omega_{0}+E_{1}\omega_{1}+E_{2}\omega_{2}
\end{aligned}
\right.
\end{equation}
where $x$ and $y$ are the state and output variable respectively. $B$, $B_{1}$, $B_{2}$, $D$, $D_{1}$, $D_{2}$ are coefficient matrices that describe the input channels of disturbances. $E$ is a coefficient matrix that is not necessarily full-ranked so as to allow for algebraic equations. In (\ref{General01}), the disturbances are of heterogeneous and isomeric nature. Specifically, $\omega_{0}$ is the unknown dynamic signal (UDS) which can represent the aerodynamic drag force. Moreover, $\omega_{1}\sim p_{\omega_{1}}(\Theta)$ is the white noise with unknown statistics represented by $\Theta$, $\omega_{2}$ is the disturbance with bounded $\mathcal{L}_{2}$-norm, $f_{\Delta}(\cdot,\cdot)$, $h_{\Delta}(\cdot)$ and $g_{\Delta}(\cdot,\cdot)$ are nonlinear mappings with parameter uncertainty denoted by $\Delta$. Note that $f_{\Delta}(\cdot,\cdot)$ may be required to satisfy certain conditions (i.e., the Lipschitz condition) so as to ensure error boundedness of the filter \cite{Lei2013}. $\Xi$ is the uncertainty term which can describe the LoE of the actuator. According to the way they inject into the system, the disturbances in (\ref{General01}) can be categorized into:
\begin{itemize}
  \item[-] Additive ones, including the UDS $\omega_{0}$, random noise $\omega_{1}$, and the norm-bounded disturbance $\omega_{2}$;
  \item[-] Multiplicative ones, including the uncertain coefficient $\Xi$ that describes LoE of the actuator;
  \item[-] Recessive ones, including the parameter uncertainty $\Delta$ and the unknown noise statistics $\Theta$.
\end{itemize}

\begin{remark}\label{remark:1}
For systems described as (\ref{General01}) where multi-source, heterogeneous, and isomeric disturbances are deeply coupled, the conventional approaches are to merge the disturbances as a lumped one and design rejection or attenuation methods based on the assumed characteristics of the lumped disturbance. For example, the lumped disturbance can be taken as an extended state and estimated via DO or ESO, or described as a norm bounded variable and attenuated via the $H_{\infty}$ techniques. However, it is difficult for the existing results that rely on lumped disturbance assumption to characterize the coupled relationships of the multi-source heterogeneous disturbances. Specifically, the causality, manifold structure, complex constraints (dynamic, static, probabilistic or mixed), interconnections (static or dynamic), and topological separability of the disturbances are partially or completely neglected in the existing estimation schemes focusing on single-type disturbances, which may result in performance degradation or filter divergence.
\end{remark}

\subsection{Generalized Observability and separability analysis}\label{sec:2-2}
Observability analysis is essential to quantifying the capability or performance bound of the state estimator. Traditionally, the observability analysis is conducted by calculating the rank of the so-called observability matrix, which reflects the internal characteristics (e.g. the system invertibility). In this sense, the conventional observability analysis is usually referred to as the {\it internal observability analysis}. For systems with multi-source, heterogenous, and isomeric disturbances, the estimation performance is not only determined by system invertibility, but also dependent on the possibility and degree of disturbance separation. Therefore, it is necessary to extend the existing observability analysis and establish a more generalized observability and separability results for systems with multi-source heterogenous disturbances.

Since the disturbance observability condition was discussed in \cite{Guo2005_NDO}, the anti-disturbance controllability analysis has been carried out in \cite{Controllability01, Controllability02} for systems with multiple disturbances. In a similar way, generalized observability and separability results can be established by explicitly figuring out which types of disturbances need to be and can be rejected/attenuated. Up to date, only few attempts have been made towards the generalized observability analysis for systems like (\ref{General01}), despite its practical importance. In \cite{Observability01}, a new observability result has been presented for the integrated MEMS/GPS system with specific constraints on the state trajectories. In \cite{Observability02}, the optimized UAV flight trajectories have been designed to improve the observability of the sensor calibration model.

As the key of generalized observability analysis, the separation of heterogeneous disturbances can be realized via either rejection or attenuation of the unfavorable signals. In both cases, a prerequisite is that the prior knowledge about the disturbance signals, including the generation, propagation and interaction mechanisms, is exploited. For example, a sinusoidal signal can be separated from the stochastic noises by utilizing the sinusoid frequency and the moment information of noise to design a so-called whitening filter \cite{LiGuo2022_APDO}.

Different from the internal observability results depending only on the coefficient matrices of system model, the generalized observability and separability conditions also depend on the actual state trajectories as well as the type, magnitude, and input channel of the disturbances. Therefore, the generalized observability of can be improved via
\begin{itemize}
  \item[i)] Deploying additional sensors or reallocating the sensing resources to collect more information;
  \item[ii)] Exploiting more prior knowledge and coupled relationships on the disturbances;
  \item[iii)] Designing excitation signal to obtain an optimal state trajectory for disturbance identification and estimation.
\end{itemize}

\section{Composite disturbance filtering: Design and optimization}\label{sec:3}
\subsection{The general framework}\label{sec:3-1}
For practical systems with multi-source heterogenous disturbances, the classical filtering schemes designed under the single-type disturbance assumption may fail to meet the task requirements. Inspired by the CHADC theory \cite{Guo2014BOOK,Guo_Cao2014survey}, the CDF framework has been established to explicitly handle the multi-source heterogenous disturbances \cite{Cao2009CDC,Guo_Cao2012,Cao_Guo2012}. In this subsection, we are dedicated to elaborate the principle and architecture of the CDF scheme.

\begin{figure*}[ht]
\centering
\includegraphics[width=0.8\textwidth]{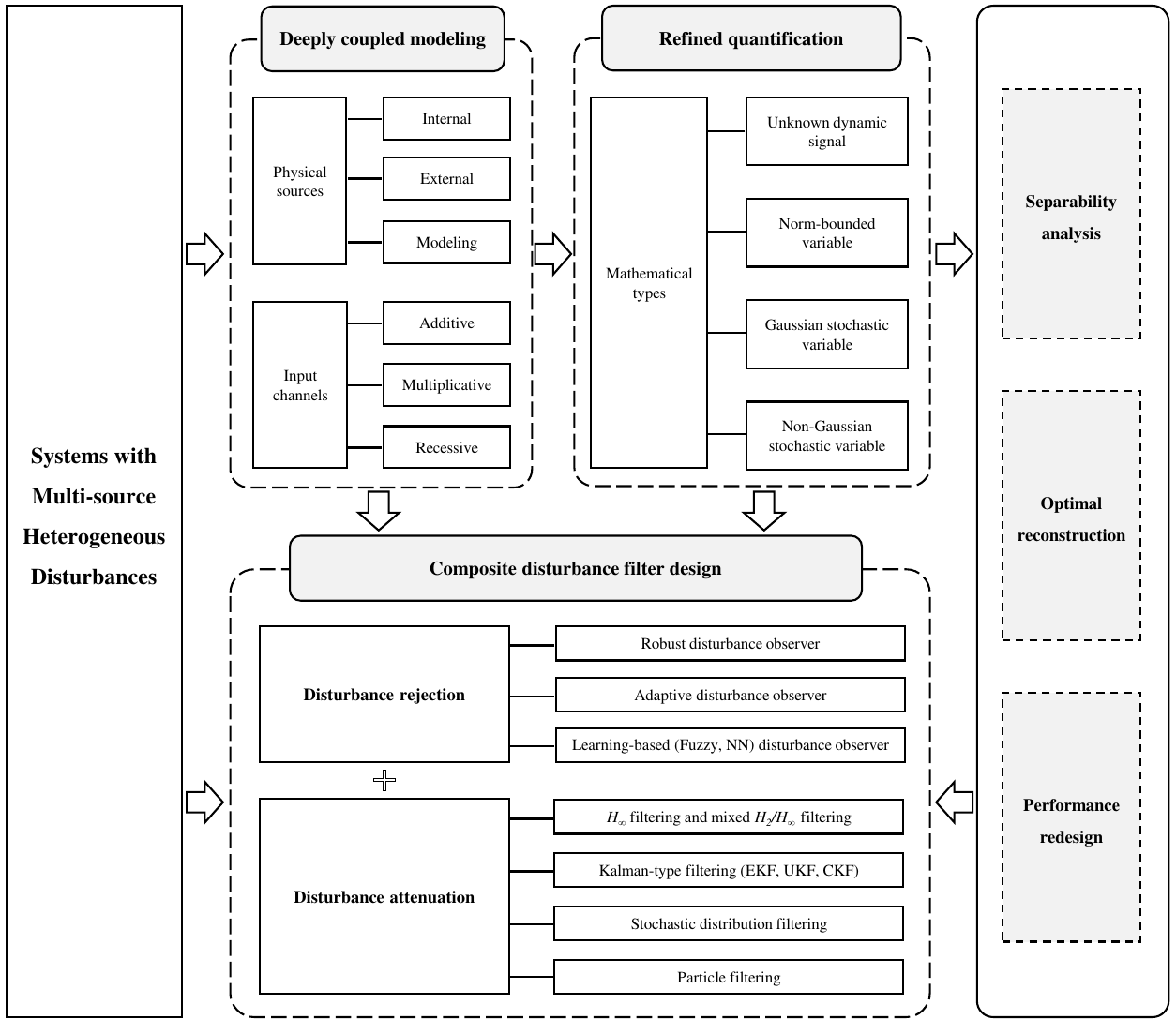}
\caption{The principle of CDF scheme.}
\label{Fig01}
\end{figure*}

The principle of the CDF is illustrated in Fig.~\ref{Fig01}. Based on deeply coupled modeling, refined quantification and separability analysis of the heterogeneous disturbances, a composite filter can be designed by combining disturbance rejection and/or attenuation strategies. Furthermore, the optimal reconstruction and performance redesign of the filter are carried out to tackle the case where the system is unobservable or different types of disturbance signals are inseparable.

\begin{figure}[ht]
\centering
\includegraphics[width=0.42\textwidth]{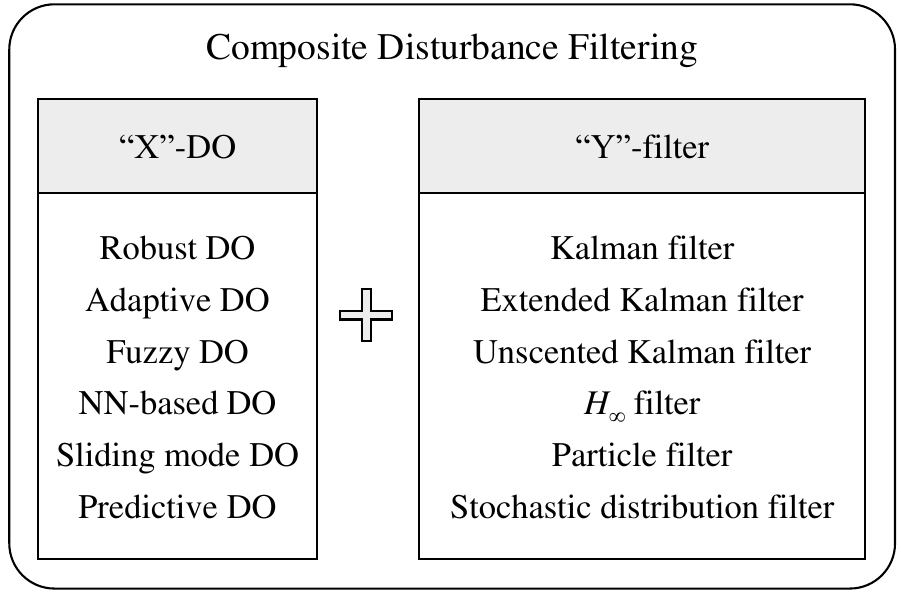}
\caption{The composite ``X-DO plus Y-Filter'' structure.}
\label{Fig04}
\end{figure}

When the heterogeneous disturbances are described as in (\ref{General01}), they can be separated and explicitly handled according to their specific characteristics. Roughly speaking, the practical disturbances can be classified into three types: the UDS, the norm-bounded variable, the Gaussian/non-Gaussian stochastic variables. With respect to the UDS, the disturbance rejection strategy can be adopted. Specifically, various DO schemes can be employed to estimate the UDSs according to their dynamic characteristics. On the other hand, the classical filtering techniques, such as $H_{\infty}$ filtering (for unknown signal with bounded $\mathcal{L}_{2}$-norm), KF (for Gaussian stochastic variable), PF and SDF (for non-Gaussian stochastic variable) can be used to quantify and optimally attenuate the effects of UDS estimation errors as well as other types of disturbances. Moreover, adaptive techniques can be adopted to handle the multiplicative or recessive disturbances without sufficient prior information. By combining the DO and the classical filtering schemes, a composite ``X-DO plus Y-Filter'' structure is constructed as illustrated in Fig.~\ref{Fig04}, which enables simultaneous rejection and attenuation of the multi-source heterogeneous disturbances.

\subsection{Composite DO+robust filter}\label{sec:3-2}
As the rudiment of CDF scheme, a composite fault diagnosis observer has been proposed in \cite{Cao2009CDC}, where the DO is combined with $H_{\infty}$ optimization techniques to separate the disturbance and fault signals. In \cite{Cao_Guo2012, Guo_Cao2012}, the composite DO+$H_{2}/H_{\infty}$ filter has been firstly developed to enhance the anti-disturbance capability for the initial alignment of INS. In the proposed scheme, the drifts of inertial sensors, modeled as the first-order Gaussian Markov process, have been estimated via the DO. Moreover, the random noises of inertial sensors (described as Gaussian stochastic variables) and the model uncertainties (described as norm-bounded variables) have been dealt with via the robust $H_{2}/H_{\infty}$ filtering technique according to a prescribed disturbance attenuation level. For the distributed state estimation problem in the simultaneous presence of UDS, norm-bounded disturbances, and false data injection (FDI) attacks, a novel composite filter has been proposed in \cite{GuoXY2022}, which is comprised of the DO for UDS estimation, the $H_{\infty}$ filter for attenuation of norm-bounded disturbances, and a detection-triggered attack estimation and rejection module for enhanced resilience against FDI attacks.

\begin{figure}[ht]
\centering
\includegraphics[width=0.42\textwidth]{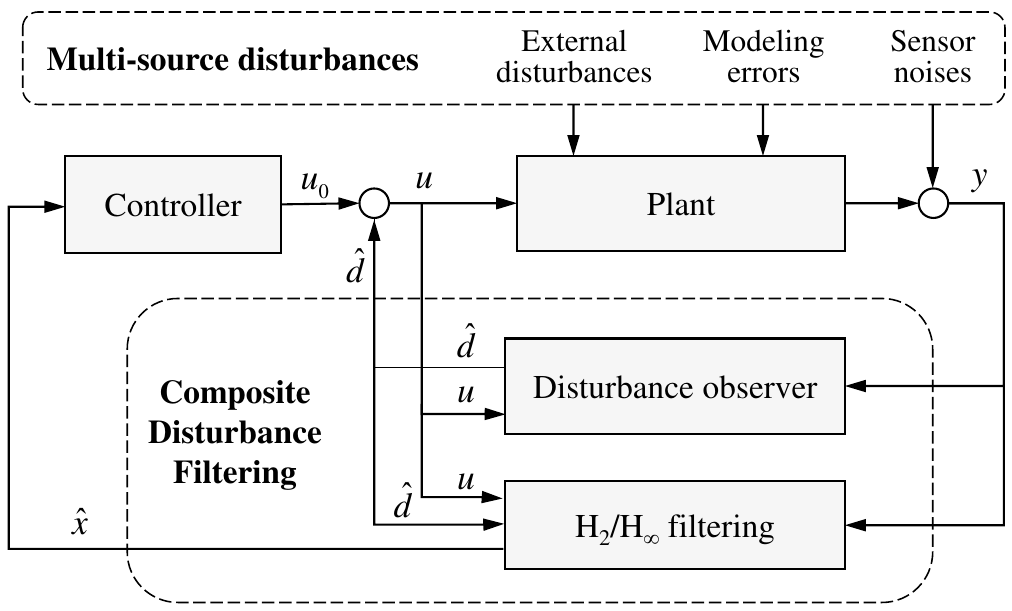}
\caption{The composite DO+$H_{2}/H_{\infty}$ filtering scheme.}
\label{Fig02}
\end{figure}

With the composite DO+robust filter, the following nonlinear system with the UDS, stochastic disturbance and norm-bounded disturbance is addressed:
\begin{equation}
\label{eqn02}
\left\{\begin{aligned}
x(k+1)&=f(x(k))+B\omega_{0}(k)+B_{1}\omega_{1}(k)+B_{2}\omega_{2}(k)\\
\omega_{0}(k+1)&=W\omega_{0}(k)+E_{1}\omega_{1}(k)+E_{2}\omega_{2}(k)\\
y(k)&=Cx(k)+D\omega_{0}(k)+D_{1}\omega_{1}(k)+D_{2}\omega_{2}(k)
\end{aligned}
\right.
\end{equation}
where $f(\cdot)$ denotes a known nonlinear mapping, $W$ and $C$ are known coefficient matrices. Note that the above system can be regarded as a simplification of (\ref{General01}), where the second equation represents the relationship among different types of disturbances. The schematic diagram of the composite DO+robust filtering scheme is illustrated in Fig.~\ref{Fig02}, where a DO is constructed to estimate the UDS in real time and the $H_{2}/H_{\infty}$ mixed optimization is utilized to attenuate the stochastic and norm-bounded disturbances. It is noted that the composite DO+robust filter enables a refined quantification of the filter errors via deeply coupled modeling. In consequence, the joint behavior of state and disturbance estimation errors can be optimized according to a prescribed performance criterion.

\subsection{Composite DO+stochastic filter}\label{sec:3-3}
\subsubsection{Composite DO+Kalman-type filter}
The composite DO+Kalman-type filtering scheme can be employed against UDS and Gaussian random noises. Specifically, the DO is utilized to estimate UDS and the KF is used to quantify and optimize the error statistics. In \cite{Du2016}, a recursive three-step information filter has been proposed for linear discrete-time systems with additive disturbances. On the basis of the traditional information filter for state estimation, an additional estimator is exploited to obtain unbiased estimate of the disturbance with minimum variance. In \cite{Du2016TIMC}, the composite DO+KF scheme has been developed for fast initial alignment of strap-down INSs. Within the composite filtering scheme, the KF is employed to estimate the horizontal misalignment angles, and the DO is designed to estimate the azimuth misalignment angle, where the steady-state output of KF is used as an input. On the other hand, when the measured data is corrupted by outliers, the measurement noise usually obeys a skew-$t$ distribution instead of the Gaussian distribution. To this end, the so-called variational Bayes skew-$t$ filtering, generalized from the KF, has recently been investigated in the literature. For example, the DO combined with the variational Bayes skew-$t$ filter is able to handle the joint effects of additive UDS and recessive statistical parameter of skew-$t$ noises \cite{Jia2022TIM}. A composite DO+learning-based KF scheme is presented in \cite{WeiYR2023} for the contact force estimation of robot manipulators. The uncertainties in the robot manipulator model and the force generative model are effectively separated by employing Gaussian process regression and variational Bayes inference techniques for statistical parameters learning.

\subsubsection{Composite DO+non-Gaussian filter}
Stochastic noises following non-Gaussian distribution widely exist in practical systems and the separation of non-Gaussian noises constitutes a key issue in signal identification and abnormity diagnosis. Nevertheless, the filter design in the presence of non-Gaussian noises has long been a challenging problem, especially when the noises are mixed with other unknown signals. Although the Bayes recursion formula has provided a generic solution to the stochastic filtering problem, it relies on an underlying assumption that the noises are independent and have identical distribution. Hence, the coupling between the UDS and stochastic disturbances needs to be addressed by developing novel filtering methods capable of separating UDSs and non-Gaussian noises.

To this end, the composite DO+non-Gaussian filtering is proposed for effective separation of UDSs and non-Gaussian noises. In \cite{Cao2014}, the fault detection problem has been addressed in the simultaneous presence of non-Gaussian noises, UDS, and norm-bounded disturbances. Specifically, the non-Gaussian PDFs have been approximated in terms of the dynamic weights of B-spline NN, and the effective separation of fault signal has been achieved via a composite DO+$H_{\infty}$ filter for simultaneous UDS rejection and norm-bounded disturbance attenuation. Note that the filtering method proposed in \cite{Cao2014} do not rely on memoryless assumption on the noise sequence due to its capability of identifying and separating the dynamic components from the non-Gaussian noises. For the system with both UDS and non-Gaussian noises, a composite DO+SDF scheme has been proposed in \cite{Yi2016TFuzzy}, where the UDS is estimated via the DO and the non-Gaussian PDF is approximated by using the fuzzy basis functions and the associated weights. As the joint PDF of the state and output variables can be tracked directly, the random noises in \cite{Yi2016TFuzzy} are not limited to be independent identically distributed ones. Moreover, composite DO plus minimum entropy filtering has been proposed in \cite{Tian2022TCYB}, where the UDS is estimated by the DO and the effect of non-Gaussian noises is attenuated based on the minimum entropy principle. Furthermore, the information-theoretic learning based CDF method has been studied in \cite{Tian2021TNNLS} for the non-Gaussian system with unknown noise PDF.

\begin{figure}[ht]
\centering
\includegraphics[width=0.42\textwidth]{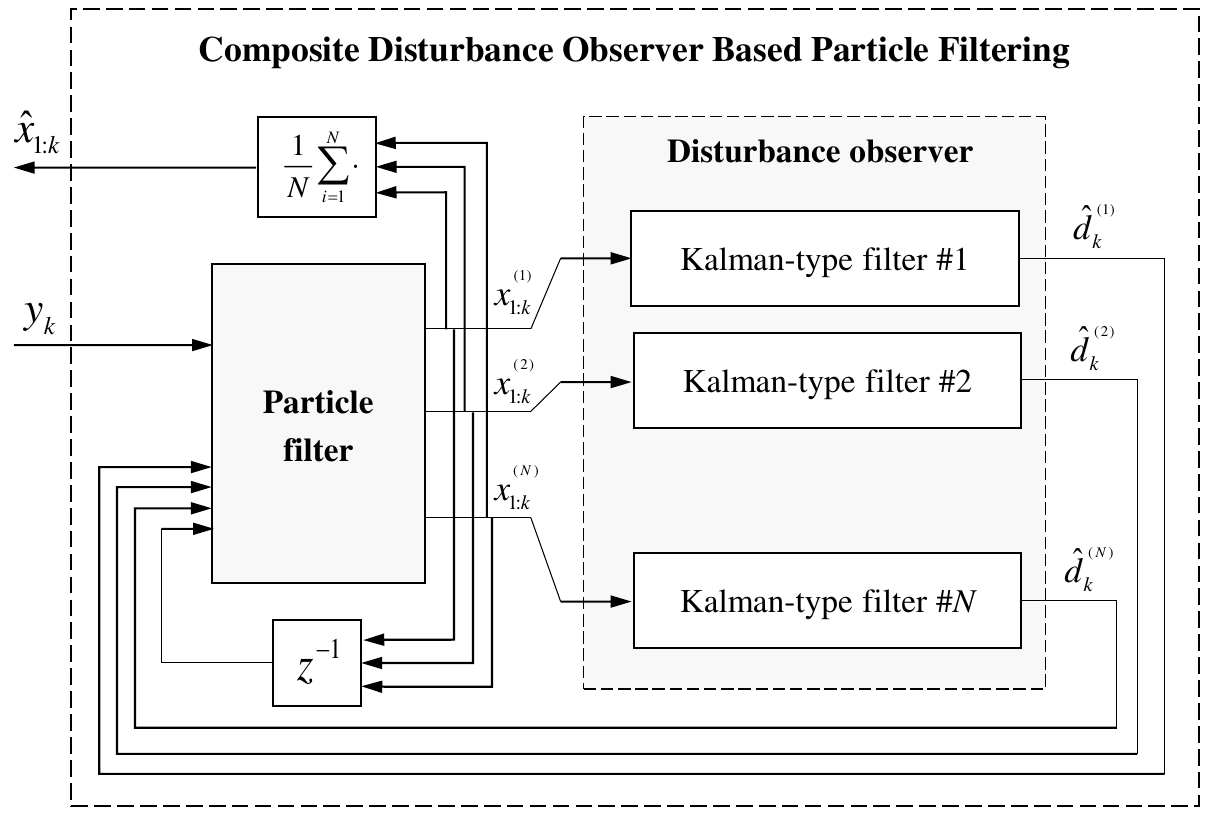}
\caption{The composite DO+PF approach.}
\label{Fig03}
\end{figure}

As a powerful tool to deal with non-Gaussianity, the PF relies on accurate statistical information for particle generation. When the system suffers from UDS and other types of model uncertainties, the so-called particle degeneracy problem will occur. That is, due to the pollution of statistical characteristics, the particles generated according to the nominal state transition model will fall into the unimportant regions of the state space, which may lead to poor approximation capability or even divergence of the filter \cite{Arulampalam2002,WLi2016PFsurvey}. To overcome this limitation, the composite DO+PF scheme has been developed within the CDF framework \cite{WLi2021TSMCA,WLi2021JFI,WLi2022TIM}. The architecture is displayed in Fig.~\ref{Fig03}. In the composite DO+PF scheme, the estimated value of the UDS is used to construct a compensation term in the state transition equation, thereby correcting the sample deviation caused by the inaccurate state transition model. A distinctive feature of this scheme is that the DO module consists of a bank of Kalman-type filters running in parallel, each associated to a particle that represents a possible realization of system states. In addition, for the more complicated case where statistical characteristics of the noises are inaccurate, a composite variational Bayes adaptive KF (VBAKF) plus PF scheme has been developed in \cite{WLi2021TSMCA}, where the VBAKF is employed for the adaptive estimation of UDS and the online identification of noise statistics. In \cite{WLi2021JFI}, a composite DO plus variational Bayes adaptive PF approach has been proposed to enhance the robustness against outliers, where the variational Bayes method is utilized to estimate the statistical parameters of heavy-tailed measurement noises. To tackle the inaccurate dynamic information of UDS, a composite student's $t$ DO plus PF scheme has been designed in \cite{WLi2022TIM}, where a student's $t$ filter is employed to track the model-inaccuracy-induced heavy-tailed posterior distribution.

\section{Composite Disturbance Filtering: Verifications and Applications}\label{sec:4}
Filter design is a key technology in the areas such as autonomous navigation and control, where the major challenge is to handle the multi-source, heterogeneous, and isomeric disturbances. So far, the CDF has been applied into a wide range of practical systems with significantly improved anti-disturbance capability (see Table.~\ref{Tab2} for a brief summary). In this section, we will introduce three typical application scenarios of the CDF.
\begin{table*}[ht]
\caption{Typical applications of the CDF}
  \label{Tab2}
  \centering
  \fontsize{10}{15}\selectfont
\begin{tabular}{C{3.1cm}IC{3.3cm}IC{4.5cm}IC{3.8cm}}
\hline
\rowcolor{gray!10} Scenario & Disturbances &  Typical method & Main advantage  \\
\hline
\thead{Initial alignment of \\ the INS}   & \thead{Inertial sensor biases \\ and modeling errors } & \thead{Composite DO+$H_{2}/H_{\infty}$ filter \cite{Guo_Cao2012}} & \thead{Quantification of disturbance \\ attenuation degree}   \\
\hline
\thead{Indoor localization of \\ rotor UAVs}   & \thead{Skewed UWB noises and \\ aerodynamic uncertainties} & \thead{Composite DO+Kalman-type filter \cite{Jia2022TIM}} & \thead{Separation of dynamic and \\ stochastic disturbances}  \\
\hline
\thead{POL aided integrated \\  attitude determination}   & \thead{Inertial sensor biases and \\ heavy-tailed POL noises} & \thead{Composite DO+particle filter \cite{WLi2021JFI}} & \thead{Reconstruction of the particle \\ generation mechanism}   \\
\hline
\end{tabular}
\end{table*}

\subsection{Initial alignment of INS: Composite DO+$H_{2}/H_{\infty}$ filter}\label{sec:4-1}
Due to the remarkable merits in terms of high autonomy, high accuracy in short time, good continuum, and insusceptibility to climate conditions, the INS has become one of the most commonly used navigation modes. The accuracy of INS depends largely on the performance of initial alignment. Hence, it makes practical sense to apply state estimation schemes to effectively solve the initial alignment problem. The state estimation model for initial alignment of the INS can be established as follows \cite{Guo_Cao2012,Cao_Guo2012}:
\begin{equation}
\label{eqn301}
\left\{\begin{aligned}
\begin{bmatrix}\dot{\phi}_{E}(t)\\ \dot{\phi}_{N}(t)\\ \dot{\phi}_{U}(t)\end{bmatrix}&=F_{0}\begin{bmatrix}\phi_{E}(t)\\ \phi_{N}(t)\\ \phi_{U}(t)\end{bmatrix}+\begin{bmatrix}\varepsilon_{E}(t)\\ \varepsilon_{N}(t)\\ \varepsilon_{U}(t)\end{bmatrix}+f_{0}(\phi)+Bd(t)\\
\begin{bmatrix}f_{E}(t)\\ f_{N}(t)\\ w_{E}(t)\end{bmatrix}&=H\begin{bmatrix}\phi_{E}(t)\\ \phi_{N}(t)\\ \phi_{U}(t)\end{bmatrix}+\begin{bmatrix}\nabla_{E}(t)\\ \nabla_{N}(t)\\ \varepsilon_{E}(t)\end{bmatrix}+g_{0}(\phi)+Dd(t)
\end{aligned}
\right.
\end{equation}
where $\phi=\begin{bmatrix}\phi_{E}(t)&\phi_{N}(t)&\phi_{U}(t)\end{bmatrix}^{T}$ are the misalignment angles; $\varepsilon=\begin{bmatrix}\varepsilon_{E}(t)&\varepsilon_{N}(t)&\varepsilon_{U}(t)\end{bmatrix}^{T}$ are the gyroscope biases expressed in the navigation frame; $\nabla_{E}(t)$ and $\nabla_{N}(t)$ are the accelerometer biases expressed in the navigation frame; $f_{E}(t)$, $f_{N}(t)$ and $w_{E}(t)$ are, respectively, the measurement outputs of east accelerometer, north accelerometer and east gyroscope; $f_{0}(\phi)$ and $g_{0}(\phi)$ are the state-dependent nonlinear terms; the disturbance term $d(t)$ is used to account for the modeling errors, which is assumed to be an unknown signal with bounded $\mathcal{L}_{2}$-norm; the coefficient matrices
$$F_{0}=\begin{bmatrix}0 & \Omega_{U} & 0 \\ -\Omega_{U} & 0 & 0 \\ 0 & 0 & 0\end{bmatrix}, \quad H=\begin{bmatrix}0 & g & 0 \\ -g & 0 & 0 \\ 0 & -\Omega_{U} & 0\end{bmatrix}$$ with $g$ and $\Omega_{U}$ being the gravity acceleration and the up component of the earth rate, respectively; $B$ and $D$ are coefficient matrices with compatible dimensions.

Denote $\omega(t)=\begin{bmatrix}\varepsilon_{E}(t) & \varepsilon_{N}(t) & \varepsilon_{U}(t) & \nabla_{E}(t) & \nabla_{N}(t)\end{bmatrix}^{T}$. The evolution of the inertial sensor biases is described by the following first-order Gaussian Markov process:
\begin{equation}
\label{eqn302}
\begin{aligned}
\dot{\omega}(t)=W\omega(t)+Ed(t)
\end{aligned}
\end{equation}
where $W=\text{diag}\{-\frac{1}{\tau_{1}},...,-\frac{1}{\tau_{5}}\}$ with $\tau_{i}~(i=1,...,5)$ being the correlation times, and $E$ is a known coefficient matrix with compatible dimension.

It is clear from (\ref{eqn301}) and (\ref{eqn302}) that the initial alignment of INS is essentially a state estimation problem with both UDS (the inertial sensor biases) and norm-bounded disturbances (the modeling error $d(t)$). In \cite{Guo_Cao2012}, a composite DO+$H_{2}/H_{\infty}$ filtering method has been proposed, where the DO is designed to compensate for the inertial sensor biases and the multi-objective $H_{2}/H_{\infty}$ optimization technique is employed to attenuate the norm-bounded disturbances. Within the proposed method, the attenuation degree of the UDS can be explicitly evaluated via
$$\gamma_{att}=\frac{\int \parallel e_{\omega}(t)\parallel_{2}^{2} \text{d}t}{\int \parallel d(t)\parallel_{2}^{2} \text{d}t},$$
which offers a quantitative criterion for the anti-disturbance capability of the filter. When comparing to the traditional UKF method, the standard deviations of the misalignment angle estimation errors have reduced by $85\%$ (levelling $x$), $92\%$ (levelling $y$), and $54\%$ (azimuth), respectively (see Table.~1 in \cite{Guo_Cao2012}).

Besides the initial alignment, the composite DO+$H_{2}/H_{\infty}$ filter is also suitable for the online calibration of micro-electro-mechanical-systems (MEMS) gyroscopes equipped on the guided projectiles. It should be mentioned that the guided projectiles are featured by high mobility and strong overload, which would induce non-negligible cross-coupling errors, scale-factor errors, and acceleration sensitivity errors in the calibration model \cite{MEMS2015,MEMS2021}. Considering the fact that these errors are of heterogenous nature, the traditional filtering schemes can be replaced by the CDF ones to achieve simultaneous rejection and attenuation.

\begin{remark}
The refined quantification of estimation errors is an important prerequisite for parameter selection and performance evaluation. Generally speaking, the error behaviors can be described from two aspects. Firstly, the asymptotic or finite-time convergence needs to be evaluated by neglecting the disturbance terms in the error dynamic equation. Secondly, the effect of stochastic or deterministic disturbances on the estimation errors needs to be characterized. For stochastic disturbances, the $H_{2}$ norm of the system or the entropy of estimation errors can be used as performance measure. For deterministic disturbances, the $H_{\infty}$ norm of the system can be adopted to characterize the boundedness of estimation errors. In the simultaneous presence of stochastic and deterministic disturbances, hybrid performance measures such as the mixed $H_{2}/H_{\infty}$ norm can be employed to provide a refined error quantification. Furthermore, there are other means of error quantification, such as Cramer-Rao Lower Bound (CRLB) computation, which quantifies the capacity of the filter in attenuating stochastic disturbances, and significant analysis, which gives a post hoc validity test for the estimation results.
\end{remark}

\subsection{Indoor localization of rotor UAVs: Composite DO+ Kalman-type filter}\label{sec:4-2}
As mentioned in Section~\ref{sec:2-1}, indoor localization of rotor UAV (by fusing UWB and UAV dynamical information) is a typical application scenario where multi-source heterogeneous disturbances need to be dealt with in the filter design. Using small-angle assumption and neglecting the higher-order terms, the localization model can be expressed as \cite{Jia2022TIM}
\begin{equation}
\label{eqn303}
\left\{\begin{aligned}
\begin{bmatrix}\ddot{x} \\ \ddot{y} \\ \ddot{z} \end{bmatrix}&=F\begin{bmatrix}\phi \\ \theta \\ \psi\end{bmatrix}+\frac{1}{m}\begin{bmatrix}0\\ 0 \\ f \end{bmatrix}+\begin{bmatrix}0\\ 0 \\ -g \end{bmatrix}+\frac{1}{m}\begin{bmatrix}d_{\omega,1}\\ d_{\omega,2} \\ d_{\omega,3} \end{bmatrix}\\
\begin{bmatrix}\ddot{\phi} \\ \ddot{\theta} \\ \ddot{\psi}\end{bmatrix}&=H\begin{bmatrix}\tau_{\phi} \\ \tau_{\theta} \\ \tau_{\psi}\end{bmatrix}+H\begin{bmatrix}-k_{d4}\dot{\phi} \\ -k_{d5}\dot{\theta} \\ -k_{d6}\dot{\psi}\end{bmatrix}\\
r_{ij}&=\|\bm{p}-\bm{p}_{i}\|-\|\bm{p}-\bm{p}_{j}\|+n_{ij}
\end{aligned}
\right.
\end{equation}
where the coefficient matrices
$$F=\begin{bmatrix}0 & g & 0 \\ -g & 0 & 0 \\ 0 & 0 & 0\end{bmatrix},\quad H=\begin{bmatrix}I^{-1}_{xx} & 0 & 0 \\ 0 & I^{-1}_{yy} & 0 \\ 0 & 0 & I^{-1}_{zz}\end{bmatrix}$$
are derived via linearization at the hovering equilibrium, $I_{xx}$, $I_{yy}$ and $I_{zz}$ denote the rotation inertia.

The main challenge of filter design arises from two aspects: 1) the aerodynamic drag forces contribute to the major uncertainty in the state transition equation; and 2) the multi-path effect and the non-line-of-sight propagation of UWB signals in the irregular indoor space may result in skewed measurement noises with inaccurate statistical parameters.

\begin{figure*}[ht]
\centering
\includegraphics[width=0.72\textwidth]{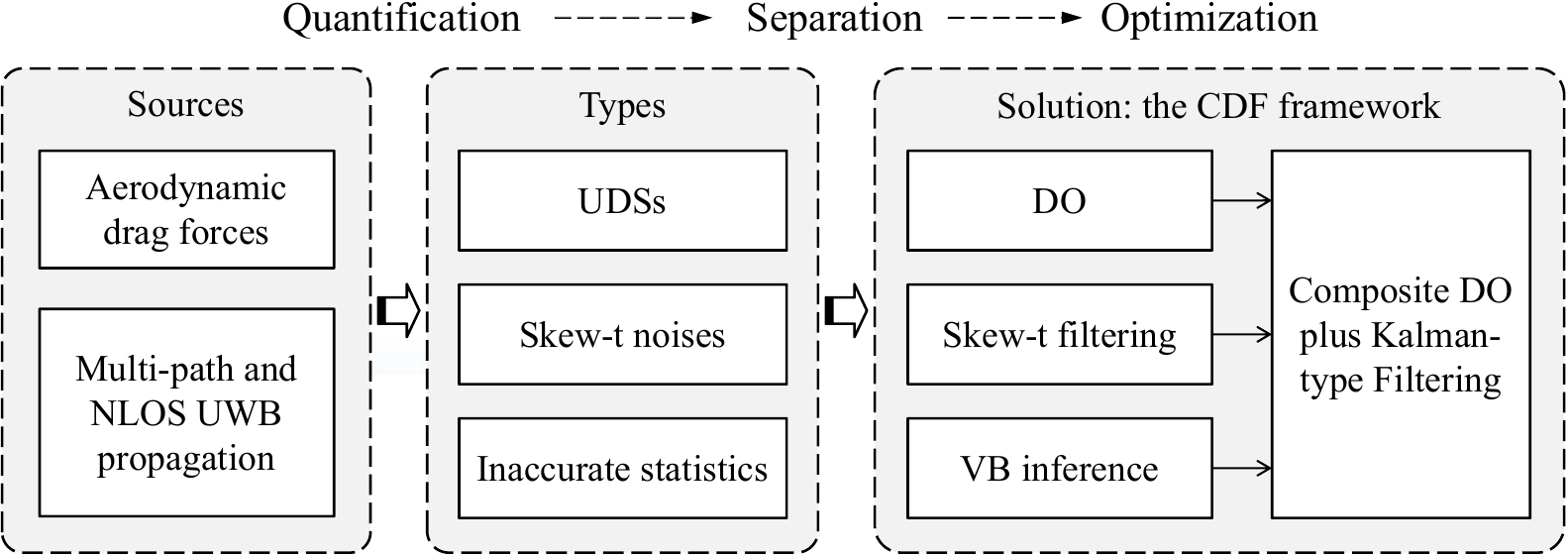}
\caption{Separation of heterogeneous disturbances in UAV localization.}
\label{Add_separation}
\end{figure*}

\begin{figure}[ht]
\centering
\includegraphics[width=0.42\textwidth]{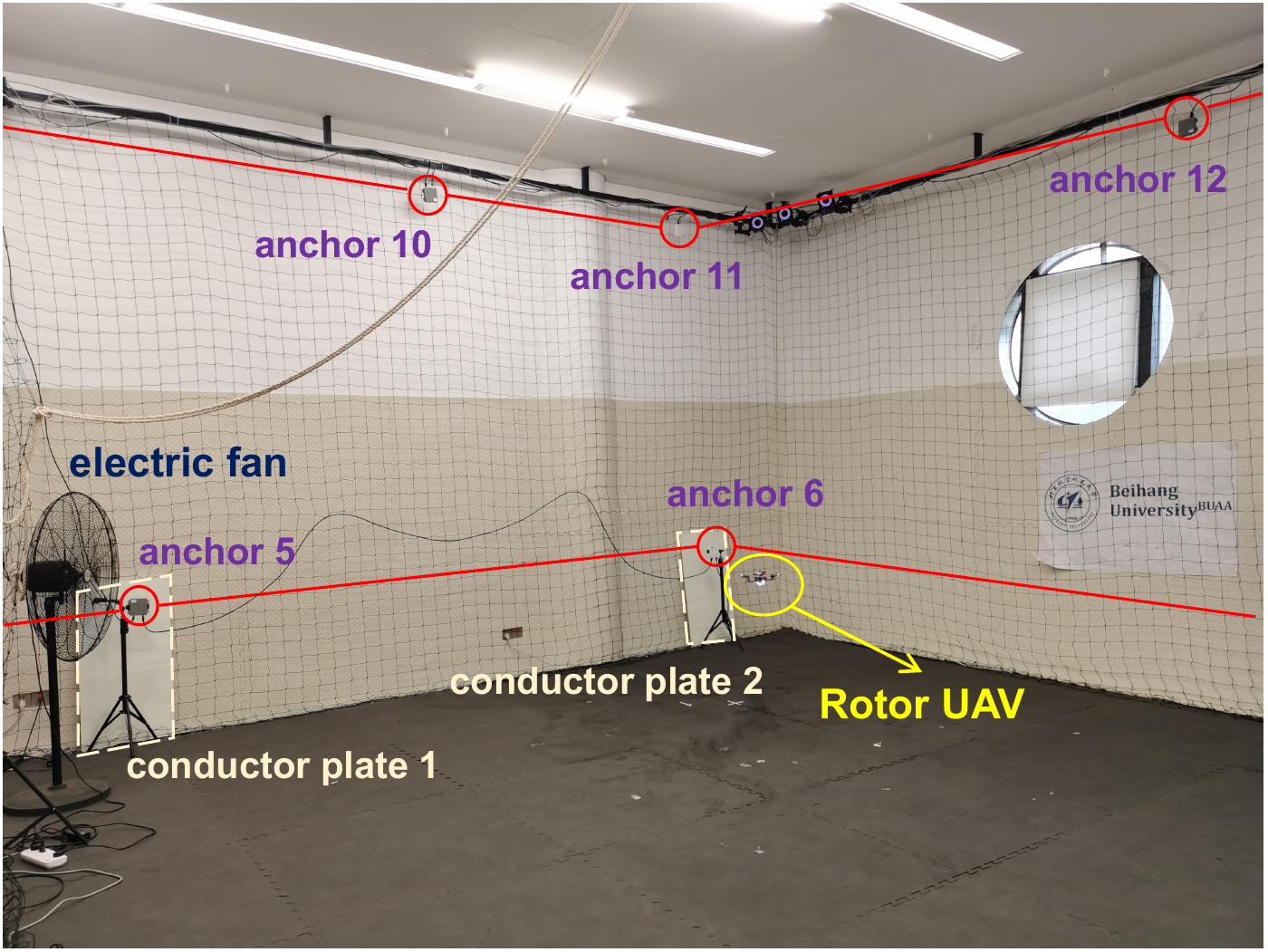}
\caption{Flight test of UAV indoor localization.}
\label{Fig05}
\end{figure}

Aiming at tackling the multi-source disturbances in UWB localization, a composite DO plus variational Bayes skew-$t$ filtering (DO+VBSTF) scheme has been developed in \cite{Jia2022TIM}, where DO is used to compensate for the uncertainties in UAV dynamics and variational Bayes skew-$t$ filter is employed to handle the skew-$t$ measurement noises by adaptively identifying the statistical parameters. With the proposed method, the dynamic uncertainties and stochastic noises can be effectively separated according to their specific characteristics (as illustrated in Fig.~\ref{Add_separation}). Experimental results on rotor UAV platform (as shown in Fig.~\ref{Fig05}) demonstrate that the proposed scheme achieves the lowest localization errors as compared with composite DO+EKF (which takes the measurement noises as Gaussian) and VBSTF (which ignores the uncertainties in the UAV dynamics). See \cite{Jia2022TIM} for detailed analysis.

\subsection{Skylight polarization aided integrated attitude determination: Composite DO+particle filter}\label{sec:4-3}
In the natural world, some insects are born with the ability to sense the polarization pattern of the skylight for autonomous navigation. Motivated by such a mechanism, the skylight polarization (POL) based navigation has emerged as an effective supplement to the traditional navigation approaches, especially in the unstructured and GPS-denied environments. Consider the following INS/POL fusion model for attitude and heading reference system (AHRS) \cite{Yang2020}
\begin{equation}
\label{eqn304}
\left\{\begin{aligned}
\dot{\theta}^{n}&=\omega_{in}^{n}\times \theta^{n}-C_{b}^{n}d_{g}^{b}+w\\
y&=\arctan\left(\frac{p^{b}[2]}{p^{b}[1]}\right)+v
\end{aligned}
\right.
\end{equation}
where the definition of symbols can be found in \cite{WLi2021JFI}.

\begin{figure}[ht]
\centering
\includegraphics[width=0.42\textwidth]{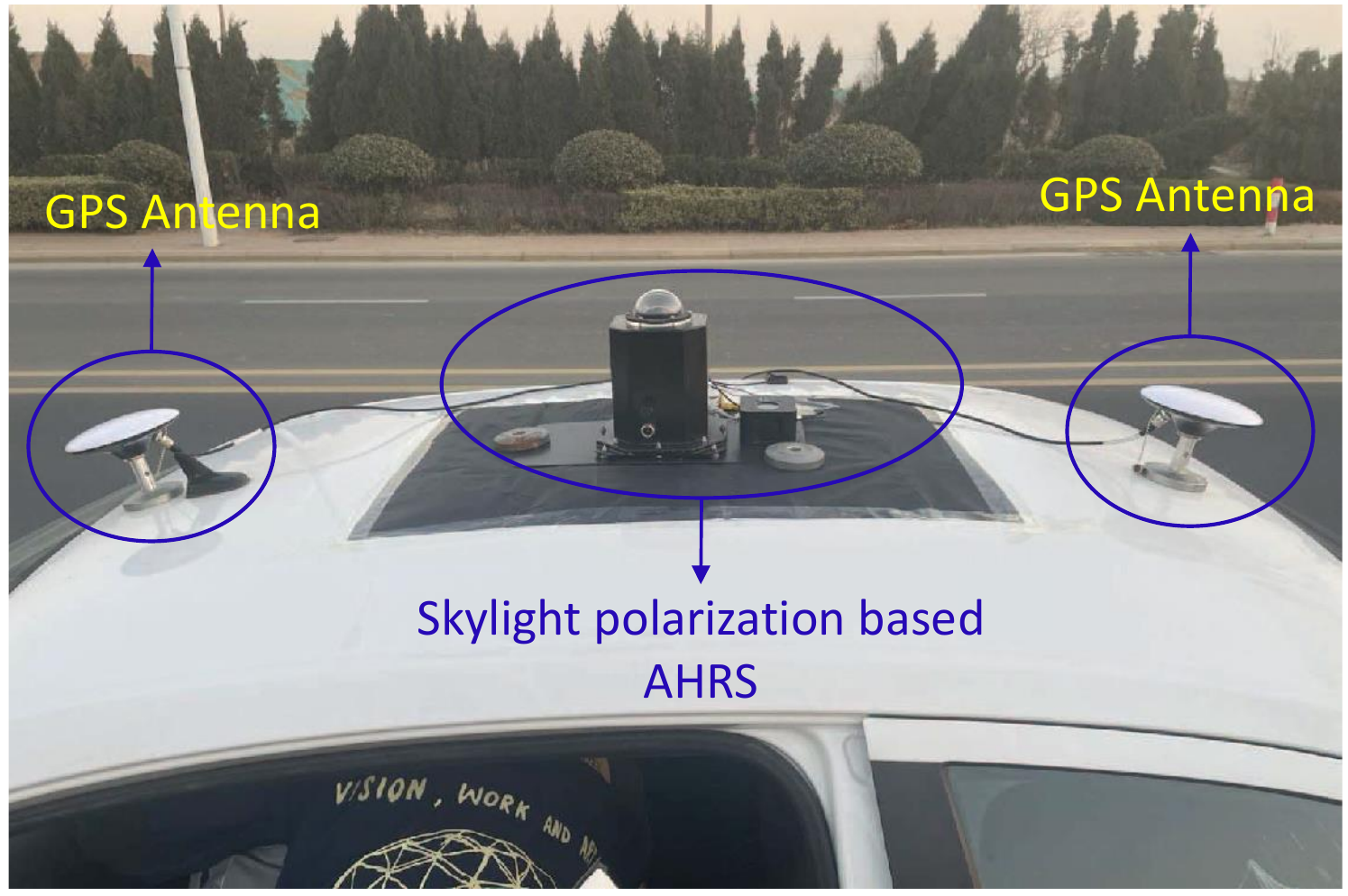}
\caption{Vehicle test.}
\label{Fig06}
\end{figure}

\begin{figure}[ht]
\centering
\includegraphics[width=0.42\textwidth]{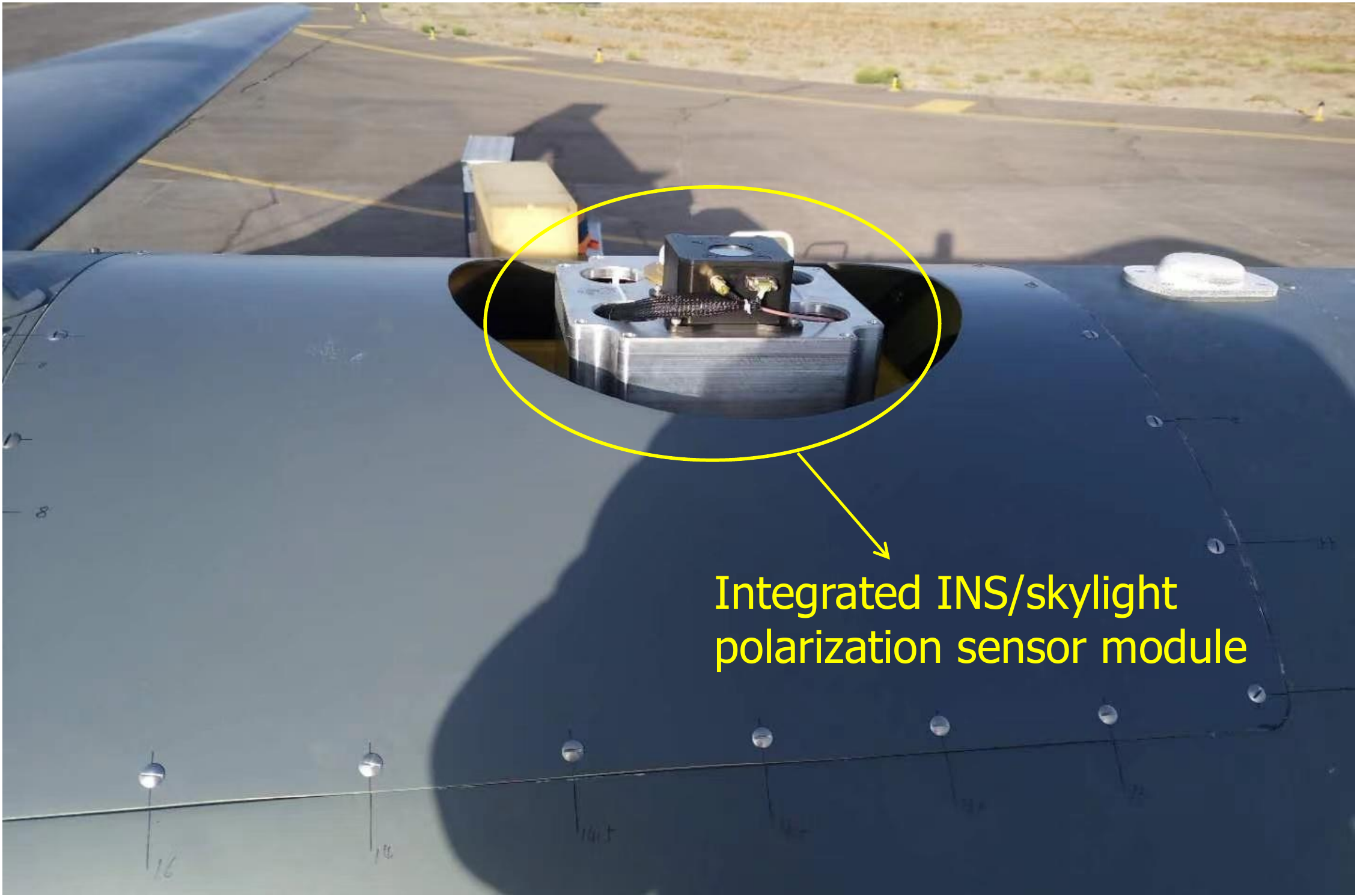}
\caption{Application to ``CH''-series UAV.}
\label{Fig07}
\end{figure}

The major challenge with INS/POL fusion lies in the coexistence of INS sensor bias and heavy-tailed POL noises arising from the occlusion of polarization sensors. In this sense, the data fusion for POL-aided AHRS can be boiled down to a filter design problem with UDS, non-Gaussian noises, and unknown noise statistics. In \cite{WLi2021JFI}, a composite DO plus variational Bayes adaptive PF (DO+VBAPF) scheme has been proposed, where a variational Bayes procedure is conducted for the online estimation of noise statistics. With the estimated UDS and noise statistics, the disturbances stemming from sensor bias and occlusion can be effectively absorbed by reconstructing the particle generation mechanism according to the updated information. The results of vehicle tests (Fig.~\ref{Fig06}) have shown that with the composite DO+VBAPF approach, the standard deviation of heading angle estimation has reduced by over $70\%$ as compared to those of the composite DO+PF and the robust PF approaches. Due to the superior anti-disturbance capability, the POL sensor together with the composite DO+VBAPF algorithm have been applied to the ``CH''-series UAV for integrated attitude determination (Fig.~\ref{Fig07}) with satisfactory accuracy and robustness.

\section{Conclusion and Outlook}\label{sec:5}
A large number of practical problems in the fields of navigation and control can be formulated as the state estimation problems under multi-source heterogenous disturbances. Nevertheless, the existing filtering schemes, such as Kalman-type filtering, $H_{\infty}$ filtering and particle filtering, have made an assumption of {\it single-type} disturbances and ignored the heterogeneity of disturbances, which may lead to unsatisfactory estimation performance in complicated task scenarios. In this paper, we have provided an overview of the CDF scheme, which is able to realize the refined state estimation by using explicit characterization, separability analysis, and a composite ``X-DO plus Y-Filter'' structure. The most significant merit of the CDF lies in the simultaneous rejection, attenuation, and absorption of multi-source heterogeneous disturbances. Different from the CHADC, the CDF does not rely on a hierarchical design procedure for disturbance compensation, but instead focuses on quantizing the behavior of disturbance estimation errors. Thanks to its capability of separating disturbance and fault signals, the CDF can be regarded as a safety-enhancing methodology for systems with persistent disturbances and abrupt faults \cite{Guo2020SSI}. Due to its refined quantification rather than brute-force treatment of the heterogeneous disturbances, the CDF can lower the sensing and computational cost in the state estimation process. Therefore, the CDF can also be viewed as a ``green''  state estimation scheme from the perspective of system resource allocation \cite{Guo2022CJA}. Furthermore, as indicated in the generalized observability theory, the state estimator can be made ``immune'' to disturbances by both passive rejection and active excitation \cite{Guo2020CJA}. Hence, the development of CDF is also a motivation of shifting the paradigm of filter design from system theory towards behavioral theory. Up to now, the effectiveness of the CDF has already been verified in a series of practical applications, including the initial alignment of INS, the UWB-based indoor localization, and the attitude determination of integrated bio-inspired/INS system.

On the basis of this paper, several potential research directions can be highlighted as follows.
\begin{itemize}
  \item[-] Conduct quantitative analysis on the generation, propagation, interaction, and influence mechanisms of multi-source heterogeneous disturbances, and establish certain refined disturbance separability results for the CDF. On this basis, enrich the ``X-DO plus Y-Filter'' framework by incorporating disturbance prediction and disturbance preview techniques \cite{LiDP2018} into the feed-forward disturbance rejection schemes, and investigating fuzzy DO, NN-based DO, and other types of adaptive DOs;
  \item[-] Design risk-awareness filters within the CDF framework to endow the information acquisition and fusion systems with immune intelligence against various active/passive disturbances including false data injection attacks \cite{Ding2019TII,Ding2021TSMCA,He2022TII}, eavesdroppers \cite{TAC2014privacy,TII2021eavesdropping,TII2021interception} and denial of services \cite{DoS2005,DoS2021,TII2021IoT}. Furthermore, improve the utilization efficiency of sensing and computational resources by developing event-triggering \cite{Event-trigger1,TII2017,Event-trigger2}, intelligent switching \cite{Scheduling2006,Scheduling2009}, distributed processing \cite{distributed01,distributed02,Dong2012TSP} and active information seeking \cite{Seeking2014,Seeking2015,Seeking2020} schemes.
  \item[-] Apply the CDF into the development of smart sensors with embedded vehicle dynamics, spatiotemporal constraints awareness, and environmental adaptability. Furthermore, conduct closed-loop uncertainty quantification based on the CDF theory to improve fault-tolerance and safety of the systems.
\end{itemize}

\end{document}